# Anisotropic Magnetoresistance and Nontrivial Spin Hall Magnetoresistance in Pt/$\alpha$-$Fe_2O_3$ Bilayers

Yang Cheng[1,*], Sisheng Yu[1,*], Adam S. Ahmed[1], Menglin Zhu[2], You Rao,[2] Maryam Ghazisaeidi,[2] Jinwoo Hwang[2], Fengyuan Yang[1]

[1]Department of Physics, The Ohio State University, Columbus, OH, USA

[2]Department of Materials Science and Engineering, The Ohio State University, Columbus, OH, 43212, USA

[*]These two authors contributed equally to this work.

Abstract

Magnetic proximity effect has only been conclusively observed in ferromagnet-based systems. We report the observation of anomalous Hall effect and angular-dependent anisotropic magnetoresistance in Pt on antiferromagnetic $\alpha$-$Fe_2O_3$(0001) epitaxial films at 10 K, which provide evidence for the magnetic proximity effect. The Néel order of $\alpha$-$Fe_2O_3$ and the induced magnetization in Pt show a unique angular dependent magnetoresistance compared with all other ferromagnetic and antiferromagnetic systems. A macrospin response model is established and can explain the antiferromagnetic spin configuration and all main magnetoresistance features in the Pt/$\alpha$-$Fe_2O_3$ bilayers.

**Introduction.** Antiferromagnets (AF) have recently generated significant excitements in spintronics due to their terahertz response, high speed, low energy cost, and abundance of materials [1-17]. Magnetic proximity effect (MPE) is an important phenomenon in spintronics with great potential for application, such as spin logic devices, modulating spin currents in graphene, and realizing high temperature quantum anomalous Hall effect in topological insulators [17-26].

To date, MPE has only been conclusively observed in nonmagnetic heavy metals (HM) on ferromagnets (FMs) [17-26]. An intriguing question is whether MPE exists in HM/AF systems. In principle, because MPE in HM/FMs originates from the surface magnetic moments, AFs with uncompensated surface moment should also be able to induce MPE. Kosub, *et al.* [27] reported that MPE might exist in Pt/$Cr_2O_3$ bilayers evidenced by the anomalous Hall effect (AHE), which is not conclusive. (see *Supplementary Materials* for more discussions). In this letter, we report the evidence of MPE in Pt/$\alpha$-$Fe_2O_3$ bilayers which exhibit both the AHE and anisotropic magnetoresistance (AMR). We explain our data by modeling the Néel order in $\alpha$-$Fe_2O_3$ and MPE-induced moment in Pt, and the competition between the spin-flop transition and the anisotropies in $\alpha$-$Fe_2O_3$.

**Experimental results.** Epitaxial $\alpha$-$Fe_2O_3$(30 nm) films are grown on $Al_2O_3$(0001) substrates using off-axis sputtering [28-30] at a substrate temperature of 500°C. After cooling down, a 2-nm Pt layer is deposited in-situ on the $\alpha$-$Fe_2O_3$ at room temperature. The quality of the films is first examined by X-ray diffraction (XRD). Figure 1(a) shows a $2\theta/\omega$ XRD scan for a Pt(2 nm)/$\alpha$-$Fe_2O_3$(30 nm) bilayer. The clear Laue oscillations (right inset) and narrow rocking curve with a full-width-at-half-maximum (FWHM) of 0.043° (left insert) indicate high crystalline quality. Figure 1(b) shows an X-ray reflectometry (XRR) scan for the 30-nm $\alpha$-$Fe_2O_3$ film, from which we extract a roughness of 0.2 nm. The smooth surface of the $\alpha$-$Fe_2O_3$ film is further confirmed by atomic force microscopy (AFM) as shown in Fig. 1(c), which gives a roughness of 0.1 nm. Figure 1(d) shows cross-sectional scanning transmission electron microscopy (STEM) images of the $\alpha$-$Fe_2O_3/Al_2O_3$ and Pt/$\alpha$-$Fe_2O_3$ interfaces which reveals clean interfaces. A few blurry clusters near the $Fe_2O_3/Al_2O_3$ interface are due to strain relaxation resulted from the ~5% lattice mismatch with the substrate. Figure 1(e) shows a magnetic hysteresis loop of a 30-nm $\alpha$-$Fe_2O_3$ film at 10 K measured by SQUID magnetometry, which only exhibits the linear diamagnetic background of $Al_2O_3$, indicating no detectable net moment (See *Supplementary Materials* for more characterizations).

The Pt(2 nm)/$\alpha$-$Fe_2O_3$(30 nm) bilayer is patterned into a 100 μm wide Hall bar. Figure 2(a) shows the Hall loops at 10 and 100 K in an out-of-plane field $\boldsymbol{H}$. Hall resistance ($R_{xy}$) generally includes the ordinary Hall effect (OHE) and anomalous Hall effect (AHE). The OHE is linearly proportional to $H$ while the AHE is typically seen in FMs and proportional to its out-of-plane magnetization [31]. In Fig. 2(a), a clear AHE signal emerges at 10 K, but disappears at 100 K.

To testify the origin of the observed AHE at 10 K and probe the spin configuration in $\alpha$-$Fe_2O_3$, we measure the magnitude of angular dependent magnetoresistance (ADMR) $\Delta\rho_{xx}/\rho_0$ for the Pt(2 nm)/$\alpha$-$Fe_2O_3$(30 nm) sample, where $\rho_0$ is the longitudinal resistivity at zero field. Figure 2(b) shows the schematics of the Hall bar with angle $\alpha$, $\beta$, and $\gamma$ defined between $\boldsymbol{H}$ and the $x$, $z$, and $z$ axes in the $xy$, $yz$, and $zx$ planes, respectively, where the current $\mathbf{I}$ is along the $x$-axis. Figure 2(c) shows the $\gamma$-scan magnetoresistance (MR) at 10 K, where a sharp peak is observed when $\boldsymbol{H} \perp$ film ($\gamma = 0°$ and 180°) at $\mu_0H$ = 1-14 T. The peak becomes narrower as $H$ increases, while the peak magnitude remains essentially the same. In Fig. 2(c), at $\mu_0H \geq 7$ T, the MR exhibits local maxima at $\gamma$ = 90° and 270°, which is a signature for the AMR [32-35]. The AMR saturates with a magnitude of ~0.01% at 7 T, which is consistent with the Hall loop at 10 K in Fig. 2(a). To explore its temperature dependence, we measure the $\gamma$-scans at 14 T for the sample at both 10 and 300 K, as shown in Fig. 2(d). The $\gamma$-scan has opposite angular dependence at 10 K as compared to 300 K (ignore the sharp peaks for now).

We next make two control samples (more discussions in the next section), a 2 nm Pt layer and a Pt(8 nm)/$\alpha$-$Fe_2O_3$(30 nm) bilayer grown on $Al_2O_3$(0001) for the same measurement at 10 K, as shown in Figs. 2(e) and 2(f). Both samples have $\sin\gamma^2$ dependence, as expected from the ordinary magnetoresistance (OMR). The OMR in Pt(2 nm)/$Al_2O_3$ is understood. The Pt(8 nm)/$\alpha$-$Fe_2O_3$ result will be discussed in the next section. Another interesting feature is that the sharp peak near $\gamma$ =180° observed in the Pt(2 nm)/$\alpha$-$Fe_2O_3$ bilayer also appears in the Pt(8 nm)/$\alpha$-$Fe_2O_3$, which cannot be explained by AMR. Below we use $\beta$- and $\alpha$-scan MR to uncover its mechanism.

Figure 3(a) shows the $\beta$-scans for the Pt/$\alpha$-$Fe_2O_3$ bilayer at 10 K. Sharp peaks are also observed for $\beta$ = 0° and 180°, but opposite to those in the $\gamma$-scans. At $\mu_0 H$ = 1 T, the $\beta$-scan has local maxima at $\beta$ = 90° and 270° with a magnitude of ~0.01%, which has been reported before in YIG/NiO/Pt [36, 37] and attributed to the Néel order $\boldsymbol{n} \perp \boldsymbol{H}$. Consequently, the $\beta$-scans show a negative spin Hall magnetoresistance (SMR) that has a 90° phase shift compared with the positive SMR in Pt/FM. As $H$ increases, the $\beta$-scans become flat and eventually have local minima at $\beta$ = 90° and 270° at 14 T due to the dominant OMR over negative SMR at high fields. By comparing the 14-T $\beta$-scans at both 10 and 300 K in Fig. 3(b) and the OMR-only 14-T $\beta$-scan of Pt(2 nm)/$Al_2O_3$ at 10 K in Fig. 3(c), it is clear that the $\beta$-scans in Fig. 3(a) are due to the competition between OMR that dominates at high fields, and the negative SMR that dominates at low fields.

Figure 3(d) shows the $\alpha$-scans for the Pt/$\alpha$-$Fe_2O_3$ bilayer at 10 K, which exhibit three notable features. First, no sharp peak is observed, but the magnitude of angular dependent MR is ~0.1%, comparable to that of the sharp peaks in the $\beta$- and $\gamma$-scans. Second, for $\mu_0 H \geq 0.3$ T, the $\alpha$-scans remain unchanged and can be well fitted by $\sin^2\alpha$ with maxima at $\alpha = 90°$ and 270°, which is a signature of negative SMR, indicating that $\boldsymbol{n}$ is perpendicular to the in-plane field due to the spin-flop transition. Third, for a small field of 0.1 T, the $\alpha$-scan deviates from $\sin^2\alpha$, suggesting that $\boldsymbol{n}$ is in a multi-domain state (below the spin-flop field). Such a small spin-flop field of <0.3 T in epitaxial $\alpha$-$Fe_2O_3$ films is 20× smaller than the ~6 T in bulk $\alpha$-$Fe_2O_3$ [38].

Figure 3(e) shows two $\alpha$-scans at 14 T, where the MR at 300 K is 50% larger than that at 10 K. As a comparison, the OMR in a Pt(2 nm)/$Al_2O_3$ control sample exhibits no $\alpha$-dependence [Fig. 3(f)]. The 300 K $\alpha$-scan in Fig. 3(e) is due to pure SMR while the 10 K data is dominated by SMR with opposite contribution from AMR, because SMR has a weak temperature dependence but AMR vanishes at 300 K. For SMR, $\Delta\rho_{xx}/\rho_0 = \theta_{\mathrm{SH}}^2 \frac{\lambda}{d}(2\lambda G \tanh^2\frac{d}{2\lambda})/(\frac{1}{\rho} + 2\lambda G \coth\frac{d}{\lambda})$, where $G$, $\theta_{\mathrm{SH}}$, $\lambda$, $d$, $\rho$ are the spin mixing conductance, spin Hall angle, spin diffusion length, thickness, and electrical resistivity of Pt, respectively [39, 40]. Using $\Delta\rho_{xx}/\rho_0 = 0.15\%$ at 300 K and 14 T in the $\alpha$-scans, $\theta_{\mathrm{SH}} = 0.086$, $\lambda = 1.2$ nm [41], $d = 2$ nm, and $\rho = 2.5 \times 10^{-7}$ Ω·m, we obtain $G = 5.5 \times 10^{15}$ $\Omega^{-1}\cdot\mathrm{m}^{-2}$, which is an order of magnitude higher than all of the other AFs [17, 42, 43].

**Discussion.** In this section, we provide likely explanations towards our experimental results. First, the Hall measurements at 10 K and 100 K shown in Fig. 2(a) suggest that our AHE signal likely arises from the MPE-induced magnetization in Pt, where the interfacial exchange interaction induces the magnetization that can be destroyed by thermal fluctuations [32, 33]. An alternative origin is the spin-Hall-induced AHE in Pt/$\alpha$-$Fe_2O_3$ [39]. However, in that case, the AHE should survive even at 300 K, because of the weak temperature dependence of spin Hall effect and the high Néel temperature ($T_N = 955$ K) of $\alpha$-$Fe_2O_3$ [44]. Second, previous studies in Pt/FM bilayers show that the existence of AMR can be used to probe the MPE. From our ADMR measurements on Pt/$\alpha$-$Fe_2O_3$, AMR is observed in the $\gamma$-scan at 10 K with a saturation magnitude of ~0.01% as shown in Fig. 2(c), which is close to the MPE-induced AMR in Pt/FM$CoFe_2O_4$ [32]. In addition, Fig. 2(d) shows the opposite angular dependencies (exclude the sharp peaks) of MR between 10 and 300 K, indicating that OMR, which has the same physical origin as OHE and has an opposite angular dependence to AMR, should dominate at 300 K. The disappearance of AMR at 300 K is expected if the AMR is induced by the MPE which is known to decrease at higher temperatures, consistent with the Hall data. Third, in our control measurements shown in Figs. 2(e) and 2(f), the Pt(8 nm)/$\alpha$-$Fe_2O_3$ bilayer exhibits no AMR signal. This rules out the possibility that the Pt magnetization is due to the formation of magnetic FePt alloy by interdiffusion, demonstrating that the induced magnetization in Pt is an interfacial effect. For the Pt(8 nm)/$\alpha$-$Fe_2O_3$, the AMR is overwhelmed by the OMR in Pt, while for the Pt(2 nm)/$\alpha$-$Fe_2O_3$, the AMR dominates OMR. Thus, the AMR in Pt(2 nm)/$\alpha$-$Fe_2O_3$ is likely due to the MPE-induced magnetization in Pt.

To explain the SMR features in the $\alpha$-, $\gamma$-, and $\beta$-scans, we use a macrospin response model (see *Supplementary Materials*) to describe AF spins in $\alpha$-$Fe_2O_3$ based on the free energy [45, 46],

$$E(\boldsymbol{n}) = H_{k1}(\boldsymbol{n}\cdot\hat{\boldsymbol{z}})^2 + H_{k2}\cos[6(\varphi_N - \delta)] + \frac{H^2}{2H_e}(\boldsymbol{h}\cdot\boldsymbol{n})^2, \quad (1)$$

where $\boldsymbol{n}$ is the unit vector of Néel order, $\varphi_N$ is the in-plane angle between $\boldsymbol{n}$ and the *x*-axis, and $\delta$ is the phase angle that defines the orientations of the easy axes. $H_{k1}$ and $H_{k2}$ are the easy-plane and in-plane easy-axis anisotropy, respectively, both of which are positive, indicating in-plane Néel order with three easy-axes 60° apart. $\boldsymbol{h}$ is the unit vector of the applied field and $H_e$ is the exchange field between the AF spins. The last term corresponds to the AF spin-flop transition, which prefers $\boldsymbol{h}\perp\boldsymbol{n}$. By minimizing $E(\boldsymbol{n})$, we can extract the Néel order in response to $\boldsymbol{H}$.

Figure 4(a) shows the schematics of $\boldsymbol{n}$ in three regimes: $\theta_H \sim 90°$, $\theta_H = 0°$, and in between, where $\theta_H$ is the angle between $\boldsymbol{H}$ and the *z*-axis, **I** || *x*-axis, and $\boldsymbol{H}$ lies in the *yz*-plane. Due to the strong easy-plane anisotropy $H_{k1}$, $\boldsymbol{n}$ always stays in-plane. The competition between

$H_{k2}$ and the spin-flop term determines $\varphi_N$. The three kinds of MR are then given by [17, 32],

$$\rho_{xx}^{\mathrm{AMR}} \propto (m_x^{\mathrm{Pt}})^2 \propto h_x^2 \propto \sin^2\theta_{\mathrm{H}},$$

$$\rho_{xx}^{\mathrm{SMR}} \propto -n_y^2 = -(1-n_x^2) \propto -\sin^2\varphi_N, \qquad (2)$$

$$\rho_{xx}^{\mathrm{OMR}} \propto -h_x^2 \propto -\sin^2\theta_{\mathrm{H}}, \text{ ($\boldsymbol{H}$ in either the $yz$- or the $zx$-plane).}$$

, where $\rho_{xx}^{\mathrm{AMR}}$ is assumed to be induced by the magnetization in Pt. To fit the MR data, we choose $H_{k1} = 2$ T, $H_{k2} = 2$ Oe, $H_e = 50$ T, and $\delta = -2.5°$ (a small non-zero $\delta$ to lift degeneracy) to first obtain the $\theta_{\mathrm{H}}$-dependence of $\varphi_N$, as shown in Fig. 4(b). As $\theta_{\mathrm{H}}$ is close to 90° [left, Fig. 4(a)], $\varphi_N$ approaches 0°, indicating that the spin-flop term dominates and $\boldsymbol{n} \perp \boldsymbol{H}$. As $\theta_{\mathrm{H}}$ rotates toward 0° [middle, Fig. 4(a)], the spin-flop term decreases and $\boldsymbol{n}$ rotates towards one of the three easy-axes. For small field like 1 T, the rotation of $\boldsymbol{n}$ is gradual. For $\mu_0 H \geq 10$ T, the in-plane component of $\boldsymbol{H}$ at $\theta_H \geq 5°$ is large enough to align $\boldsymbol{n}$ along the $x$-axis ($\varphi_N = 0°$), and at $\theta_{\mathrm{H}} < 5°$, $\varphi_N$ increases dramatically. At $\theta_{\mathrm{H}} = 0°$ [right, Fig. 4(a)], the spin-flop term is 0 and $\boldsymbol{n}$ has equal probability to align along any of the three easy axes, forming multi-domains.

This mechanism can simultaneously explain the sharp peaks in the $\gamma$- and $\beta$-scans, the small negative SMR in the $\beta$-scans, and the large negative SMR in the $\alpha$-scans. For the $\gamma$-scans away from 0° or 180°, the in-plane component of $\boldsymbol{H}$ is large enough to induce the spin-flop transition and the $\alpha$-$Fe_2O_3$ film is a single domain with $\boldsymbol{n} \parallel y$-axis. As $\gamma$ approaches 0° or 180°, the in-plane component of $\boldsymbol{H}$ drops below the spin-flop field and the $\alpha$-$Fe_2O_3$ film forms multi-domains with a significant $n_x$, resulting in a sharp increase in SMR. Meanwhile, in the $\beta$-scans, as $\beta$ approaches 0° or 180°, the $\alpha$-$Fe_2O_3$ film enters the multi-domain regime and $\boldsymbol{n}$ changes from $\boldsymbol{n} \parallel x$-axis to having a significant $n_x$, resulting in a sudden decrease in SMR. The peaks are sharper at higher fields because of the constant in-plane spin-flop field and sharper change of $\varphi_N$ [see Fig. 4(b)]

In Fig. 3(c), at $\mu_0 H = 0.1$ T, the $\alpha$-scan deviates from $\sin^2\alpha$, indicating the AF is in multi-domains.

An in-plane $\mu_0 H \geq 0.3$ T overcomes $H_{k2}$ and induces the spin-flop transition to form a single domain, which corresponds to 1.2° off the $z$-axis at 14 T. The magnitude of AMR in the $\alpha$-scans is similar to that in the $\gamma$-scans, which is ~10× smaller than the SMR with opposite angular dependence. Using $H_{k1} = 2$ T, $H_{k2} = 2$ Oe, $H_e = 50$ T, and $\delta = -2.5°$, we show the fitting of the $\gamma$- and $\beta$-scans in Fig. 4(c) and 4(d), respectively. The fitting reproduces all the key features discussed above, including the whole angular and field range of the ADMR.

**Conclusion.** Our experimental results provide evidence for MPE in the Pt/$\alpha$-$Fe_2O_3$ bilayers, which most likely arises from the uncompensated surface spins of $\alpha$-$Fe_2O_3$ [27, 47]. In our model, we assume that the AMR is induced by the MPE and the MPE-induced magnetization in Pt is parallel to $\boldsymbol{H}$, with fits our experimental data well. Our modeling shows that the Néel order is not parallel to induced magnetization in Pt, which is different from the FM-induced MPE, where the FM magnetization and the MPE-induced moment in Pt both follow $\boldsymbol{H}$. We note that our results cannot fully rule out other possible mechanisms for MPE, such as local ferromagnetism induced by nonuniformities or defects at the interface. To better understand the mechanism, more theoretical investigations and direct detection (e.g., using X-ray magnetic circular dichroism) are needed.

This work was supported primarily the U.S. Department of Energy (DOE) under Grant No. DE-SC0001304 (YC, SSY, and FYY), and partially supported by the Center for Emergent Materials, an NSF MRSEC, under Grant No. DMR-1420451 (ASA, MLZ, and JH).

**Figure Captions:**

**Figure 1.** (a) XRD $2\theta/\omega$ scan of a Pt(2 nm)/$\alpha$-$Fe_2O_3$(30 nm) bilayer on $Al_2O_3$(0001). Right inset: high-resolution scan of the $\alpha$-$Fe_2O_3$ (0006) peak with Laue oscillations. Left inset: XRD rocking curve of the $\alpha$-$Fe_2O_3$ (0006) peak. (b) XRR scan of a 30 nm $\alpha$-$Fe_2O_3$ film on $Al_2O_3$(0001). (c) AFM image of an α-$Fe_2O_3$(30 nm) with a roughness of 0.1 nm. (d) STEM images of the $\alpha$-$Fe_2O_3$/$Al_2O_3$ and Pt/$\alpha$-$Fe_2O_3$ interfaces viewed along $(1\bar{2}10)$ and $(5\bar{4}10)$, respectively. (e) Magnetic hysteresis loop of a 30 nm $\alpha$-$Fe_2O_3$ film taken at 10 K with no detectable magnetization.

**Figure 2.** (a) Hall resistance for a Pt(2 nm)/$\alpha$-$Fe_2O_3$(30 nm) bilayer at 10 and 100 K. (b) Schematics of $\alpha$ (*xy*-plane), $\gamma$ (*zx*-plane), and $\beta$ (*yz*-plane) angular dependence measurements. (c) $\gamma$-scans of a Pt(2 nm)/$\alpha$-$Fe_2O_3$(30 nm) sample at 10 K an various magnetic fields, which show a sharp peak at out-of-plane field (0°, and 180°) and a broad peak at in-plane field (90° and 270°). Curves are shifted vertically for clarity. (d) $\gamma$-scans at 10 and 300 K for Pt(2 nm)/$\alpha$-$Fe_2O_3$(30 nm) at 14 T. Control experiments of $\gamma$-scans for (e) a Pt(8 nm)/$\alpha$-$Fe_2O_3$(30nm) bilayer and (f) a Pt(2 nm) on $Al_2O_3$(0001) taken at 14 T and 10 K, where OMR dominates the angular dependence. The red curves in (e) and (f) are cosine fits.

**Figure 3.** (a) $\beta$-scans of a Pt(2 nm)/$\alpha$-$Fe_2O_3$(30 nm) bilayer at 10 K and various fields. (b) Comparison of the $\beta$-scans between 10 and 300 K at 14 T. (c) $\beta$-scan of a control sample Pt(2 nm) on $Al_2O_3$(0001) at 14 T and 10 K. (d) $\alpha$-scans of a Pt(2 nm)/$\alpha$-$Fe_2O_3$(30 nm) bilayer at 10 K and various fields. (e) Comparison of the $\alpha$-scans between 10 and 300 K at 14 T. The solid curves in (c), (d), and (e) are cosine fits. (f) $\alpha$-scan of a control sample Pt(2 nm) on $Al_2O_3$ (0001) at 14 T and 10 K. Curves are shifted vertically for clarity.

**Figure 4.** (a) Schematics of AF spin configurations as an applied field rotates from in-plane

towards out-of-plane in the $yz$-plane (for $\beta$ scans), where the green lines illustrate the three in-plane easy axes of $\alpha$-$Fe_2O_3$. (b) Simulation of $\varphi_N$, the angle between the Néel order $\boldsymbol{n}$ and the $x$-axis, at different field angle $\theta_H$. Fitting of (c) $\gamma$-scans and (d) $\beta$-scans at 10 K for various fields, where the solid curves are fits to the experimental data. Curves are shifted vertically for clarity.

**References:**

1. W. Zhang, M. B. Jungfleisch, W. J. Jiang, J. E. Pearson, A. Hoffmann, F. Freimuth and Y. Mokrousov, "Spin Hall Effects in Metallic Antiferromagnets," *Phys. Rev. Lett.* **113**, 196602 (2014).
2. T. Jungwirth, X. Marti, P. Wadley and J. Wunderlich, "Antiferromagnetic spintronics," *Nat. Nanotechnol.* **11**, 231 (2016).
3. P. Wadley, B. Howells, J. Zelezny, C. Andrews, V. Hills, R. P. Campion, V. Novak, K. Olejnik, F. Maccherozzi, S. S. Dhesi, S. Y. Martin, T. Wagner, J. Wunderlich, F. Freimuth, Y. Mokrousov, J. Kunes, J. S. Chauhan, M. J. Grzybowski, A. W. Rushforth, K. W. Edmonds, B. L. Gallagher and T. Jungwirth, "Electrical switching of an antiferromagnet," *Science* **351**, 587 (2016).
4. A. H. MacDonald and M. Tsoi, "Antiferromagnetic metal spintronics," *Phil. Trans. Roy. Soc. A-Math. Phys. Eng. Sci.* **369**, 3098 (2011).
5. S. Urazhdin and N. Anthony, "Effect of polarized current on the magnetic state of an antiferromagnet," *Phys. Rev. Lett.* **99**, 046602 (2007).
6. R. Cheng, J. Xiao, Q. Niu and A. Brataas, "Spin Pumping and Spin-Transfer Torques in Antiferromagnets," *Phys. Rev. Lett.* **113**, 057601 (2014).
7. X. Marti, I. Fina, C. Frontera, J. Liu, P. Wadley, Q. He, R. J. Paull, J. D. Clarkson, J. Kudrnovsky, I. Turek, J. Kunes, D. Yi, J. H. Chu, C. T. Nelson, L. You, E. Arenholz, S. Salahuddin, J. Fontcuberta, T. Jungwirth and R. Ramesh, "Room-temperature antiferromagnetic memory resistor," *Nat. Mater.* **13**, 367 (2014).
8. S. Nakatsuji, N. Kiyohara and T. Higo, "Large anomalous Hall effect in a non-collinear antiferromagnet at room temperature," *Nature* **527**, 212 (2015).
9. H. L. Wang, C. H. Du, P. C. Hammel and F. Y. Yang, "Antiferromagnonic Spin Transport from $Y_3Fe_5O_{12}$ into NiO," *Phys. Rev. Lett.* **113**, 097202 (2014).
10. T. Satoh, R. Iida, T. Higuchi, M. Fiebig and T. Shimura, "Writing and reading of an arbitrary optical polarization state in an antiferromagnet," *Nat. Photonics* **9**, 25 (2015).
11. H. L. Wang, C. H. Du, P. C. Hammel and F. Y. Yang, "Spin transport in antiferromagnetic insulators mediated by magnetic correlations," *Phys. Rev. B* **91**, 220410(R) (2015).
12. R. Cheng, D. Xiao and A. Brataas, "Terahertz Antiferromagnetic Spin Hall Nano-Oscillator," *Phys. Rev. Lett.* **116**, 207603 (2016).
13. A. Prakash, J. Brangham, F. Y. Yang and J. P. Heremans, "Spin Seebeck effect through antiferromagnetic NiO," *Phys. Rev. B* **94**, 014427 (2016).
14. T. Kampfrath, A. Sell, G. Klatt, A. Pashkin, S. Mahrlein, T. Dekorsy, M. Wolf, M. Fiebig, A. Leitenstorfer and R. Huber, "Coherent terahertz control of antiferromagnetic spin waves," *Nat. Photonics* **5**, 31 (2011).
15. S. Seki, T. Ideue, M. Kubota, Y. Kozuka, R. Takagi, M. Nakamura, Y. Kaneko, M. Kawasaki and Y. Tokura, "Thermal Generation of Spin Current in an Antiferromagnet," *Phys. Rev. Lett.* **115**, 266601 (2015).
16. X. Z. Chen, R. Zarzuela, J. Zhang, C. Song, X. F. Zhou, G. Y. Shi, F. Li, H. A. Zhou, W. J. Jiang, F. Pan and Y. Tserkovnyak, "Antidamping-Torque-Induced Switching in Biaxial Antiferromagnetic Insulators," *Phys. Rev. Lett.* **120**, 207204 (2018).
17. L. Baldrati, A. Ross, T. Niizeki, C. Schneider, R. Ramos, J. Cramer, O. Gomonay, M. Filianina, T. Savchenko, D. Heinze, A. Kleibert, E. Saitoh, J. Sinova and M. Klaui, "Full angular dependence of the spin Hall and ordinary magnetoresistance in epitaxial antiferromagnetic NiO(001)/Pt thin films," *Phys. Rev. B* **98**, 024422 (2018).
18. Z. L. Jiang, C. Z. Chang, C. Tang, P. Wei, J. S. Moodera and J. Shi, "Independent Tuning of Electronic Properties and Induced Ferromagnetism in Topological Insulators with Heterostructure Approach," *Nano Letters* **15**, 5835 (2015).
19. I. Vobornik, U. Manju, J. Fujii, F. Borgatti, P. Torelli, D. Krizmancic, Y. S. Hor, R. J. Cava and G. Panaccione, "Magnetic Proximity Effect as a Pathway to Spintronic Applications of Topological Insulators," *Nano Letters* **11**, 4079 (2011).
20. S. Y. Huang, X. Fan, D. Qu, Y. P. Chen, W. G. Wang, J. Wu, T. Y. Chen, J. Q. Xiao and C. L. Chien, "Transport Magnetic Proximity Effects in Platinum," *Phys. Rev. Lett.* **109**, 107204 (2012).
21. H. Nakayama, M. Althammer, Y. T. Chen, K. Uchida, Y. Kajiwara, D. Kikuchi, T. Ohtani, S. Geprags, M. Opel, S. Takahashi, R. Gross, G. E. W. Bauer, S. T. B. Goennenwein and E. Saitoh, "Spin Hall Magnetoresistance Induced by a Nonequilibrium Proximity Effect," *Phys. Rev. Lett.* **110**, 206601 (2013).
22. M. D. Li, C. Z. Chang, B. J. Kirby, M. E. Jamer, W. P. Cui, L. J. Wu, P. Wei, Y. M. Zhu, D. Heiman, J. Li and J. S. Moodera, "Proximity-Driven Enhanced Magnetic Order at Ferromagnetic-Insulator-Magnetic-Topological-Insulator Interface," *Phys. Rev. Lett.* **115**, 087201 (2015).
23. Z. Wang, C. Tang, R. Sachs, Y. Barlas and J. Shi, "Proximity-Induced Ferromagnetism in Graphene Revealed by the Anomalous Hall Effect," *Phys. Rev. Lett.* **114**, 016603 (2015).
24. F. Katmis, V. Lauter, F. S. Nogueira, B. A. Assaf, M. E. Jamer, P. Wei, B. Satpati, J. W. Freeland, I. Eremin, D. Heiman, P. Jarillo-Herrero and J. S. Moodera, "A high-temperature ferromagnetic topological insulating phase by proximity coupling," *Nature* **533**, 513 (2016).
25. F. Hellman, A. Hoffmann, Y. Tserkovnyak, G. S. D. Beach, E. E. Fullerton, C. Leighton, A. H. MacDonald, D. C. Ralph, D. A. Arena, H. A. Durr, P. Fischer, J. Grollier, J. P. Heremans, T. Jungwirth, A. V. Kimel, B. Koopmans, I. N. Krivorotov, S. J. May, A. K. Petford-Long, J. M. Rondinelli, N. Samarth, I. K. Schuller, A. N. Slavin, M. D. Stiles, O. Tchernyshyov, A. Thiaville and B. L. Zink, "Interface-induced phenomena in magnetism," *Rev. Mod. Phys.* **89**, 025006 (2017).
26. S. Singh, J. Katoch, T. C. Zhu, K. Y. Meng, T. Y. Liu, J. T. Brangham, F. Y. Yang, M. E. Flatte and R. K. Kawakami, "Strong Modulation of Spin Currents in Bilayer Graphene by Static and Fluctuating Proximity Exchange Fields," *Phys. Rev. Lett.* **118**, 187201 (2017).
27. T. Kosub, M. Kopte, F. Radu, O. G. Schmidt and D. Makarov, "All-Electric Access to the Magnetic-Field-Invariant Magnetization of Antiferromagnets," *Phys. Rev. Lett.* **115**, 097201 (2015).
28. F. Y. Yang and P. C. Hammel, "Topical review: FMR-Driven Spin Pumping in $Y_3Fe_5O_{12}$-Based Structures," *J. Phys. D: Appl. Phys.* **51**, 253001 (2018).
29. A. J. Hauser, R. E. A. Williams, R. A. Ricciardo, A. Genc, M. Dixit, J. M. Lucy, P. M. Woodward, H. L. Fraser and F. Y. Yang, "Unlocking the potential of half-metallic $Sr_2FeMoO_6$ films through controlled stoichiometry and double-perovskite ordering," *Phys. Rev. B* **83**, 014407 (2011).
30. B. Peters, A. Alfonsov, C. G. F. Blum, S. J. Hageman, P. M. Woodward, S. Wurmehl, B. Büchner and F. Y. Yang, "Epitaxial films of Heusler compound $Co_2FeAl_{0.5}Si_{0.5}$ with high crystalline quality grown by off-axis sputtering," *Appl. Phys. Lett.* **103**, 162404 (2013).

31. N. Nagaosa, J. Sinova, S. Onoda, A. H. MacDonald and N. P. Ong, "Anomalous Hall effect," *Rev. Mod. Phys.* **82**, 1539 (2010).
32. W. Amamou, I. V. Pinchuk, A. H. Trout, R. E. A. Williams, N. Antolin, A. Goad, D. J. O'Hara, A. S. Ahmed, W. Windl, D. W. McComb and R. K. Kawakami, "Magnetic proximity effect in Pt/$CoFe_2O_4$ bilayers," *Phys. Rev. Mater.* **2**, 011401 (2018).
33. X. Zhou, L. Ma, Z. Shi, W. J. Fan, J.-G. Zheng, R. F. L. Evans and S. M. Zhou, "Magnetotransport in metal/insulating-ferromagnet heterostructures: Spin Hall magnetoresistance or magnetic proximity effect," *Phys. Rev. B* **92**, 060402 (2015).
34. X. Liang, G. Y. Shi, L. J. Deng, F. Huang, J. Qin, T. T. Tang, C. T. Wang, B. Peng, C. Song and L. Bi, "Magnetic Proximity Effect and Anomalous Hall Effect in Pt/$Y_3Fe_{5-x}Al_xO_{12}$ Heterostructures," *Phys. Rev. Appl.* **10**, 024051 (2018).
35. T. Shang, Q. F. Zhan, H. L. Yang, Z. H. Zuo, Y. L. Xie, L. P. Liu, S. L. Zhang, Y. Zhang, H. H. Li, B. M. Wang, Y. H. Wu, S. Zhang and R.-W. Li, "Effect of NiO inserted layer on spin-Hall magnetoresistance in Pt/NiO/YIG heterostructures," *Appl. Phys. Lett.* **109**, 032410 (2016).
36. D. Z. Hou, Z. Y. Qiu, J. Barker, K. Sato, K. Yamamoto, S. Vélez, J. M. Gomez-Perez, L. E. Hueso, F. Casanova and E. Saitoh, "Tunable Sign Change of Spin Hall Magnetoresistance in Pt/NiO/YIG Structures," *Phys. Rev. Lett.* **118**, 147202 (2017).
37. W. W. Lin and C. L. Chien, "Electrical Detection of Spin Backflow from an Antiferromagnetic Insulator/$Y_3Fe_5O_{12}$ Interface," *Phys. Rev. Lett.* **118**, 067202 (2017).
38. R. Lebrun, A. Ross, S. A. Bender, A. Qaiumzadeh, L. Baldrati, J. Cramer, A. Brataas, R. A. Duine and M. Kläui, "Tunable long-distance spin transport in a crystalline antiferromagnetic iron oxide," *Nature* **561**, 222 (2018).
39. Y. T. Chen, S. Takahashi, H. Nakayama, M. Althammer, S. T. B. Goennenwein, E. Saitoh and G. E. W. Bauer, "Theory of spin Hall magnetoresistance," *Phys. Rev. B* **87**, 144411 (2013).
40. H. L. Wang, C. H. Du, P. C. Hammel and F. Y. Yang, "Comparative determination of $Y_3Fe_5O_{12}$/Pt interfacial spin mixing conductance by spin-Hall magnetoresistance and spin pumping," *Appl. Phys. Lett.* **110**, 062402 (2017).
41. W. Zhang, V. Vlaminck, J. E. Pearson, R. Divan, S. D. Bader and A. Hoffmann, "Determination of the Pt spin diffusion length by spin-pumping and spin Hall effect," *Appl. Phys. Lett.* **103**, 242414 (2013).
42. Y. Ji, J. Miao, K. K. Meng, Z. Y. Ren, B. W. Dong, X. G. Xu, Y. Wu and Y. Jiang, "Spin Hall magnetoresistance in an antiferromagnetic magnetoelectric $Cr_2O_3$/heavy-metal W heterostructure," *Appl. Phys. Lett.* **110**, 262401 (2017).
43. J. H. Han, C. Song, F. Li, Y. Y. Wang, G. Y. Wang, Q. H. Yang and F. Pan, "Antiferromagnet-controlled spin current transport in $SrMnO_3$/Pt hybrids," *Phys. Rev. B* **90**, 144431 (2014).
44. F. Bødker, M. F. Hansen, C. B. Koch, K. Lefmann and S. Mørup, "Magnetic properties of hematite nanoparticles," *Phys. Rev. B* **61**, 6826 (2000).
45. Y. Cheng, R. Zarzuela, J. T. Brangham, A. J. Lee, S. White, P. C. Hammel, Y. Tserkovnyak and F. Y. Yang, "Nonsinusoidal angular dependence of FMR-driven spin current across an antiferromagnet in $Y_3Fe_5O_{12}$/NiO/Pt trilayers," *Phys. Rev. B* **99**, 060405 (2019).
46. J. Fischer, O. Gomonay, R. Schlitz, K. Ganzhorn, N. Vlietstra, M. Althammer, H. Huebl, M. Opel, R. Gross, S. T. B. Goennenwein and S. Geprägs, "Spin Hall magnetoresistance in antiferromagnet/heavy-metal heterostructures," *Phys. Rev. B* **97**, 014417 (2018).

47. X. He, Y. Wang, N. Wu, A. N. Caruso, E. Vescovo, K. D. Belashchenko, P. A. Dowben and C. Binek, "Robust isothermal electric control of exchange bias at room temperature," *Nat. Mater.* **9**, 579 (2010).

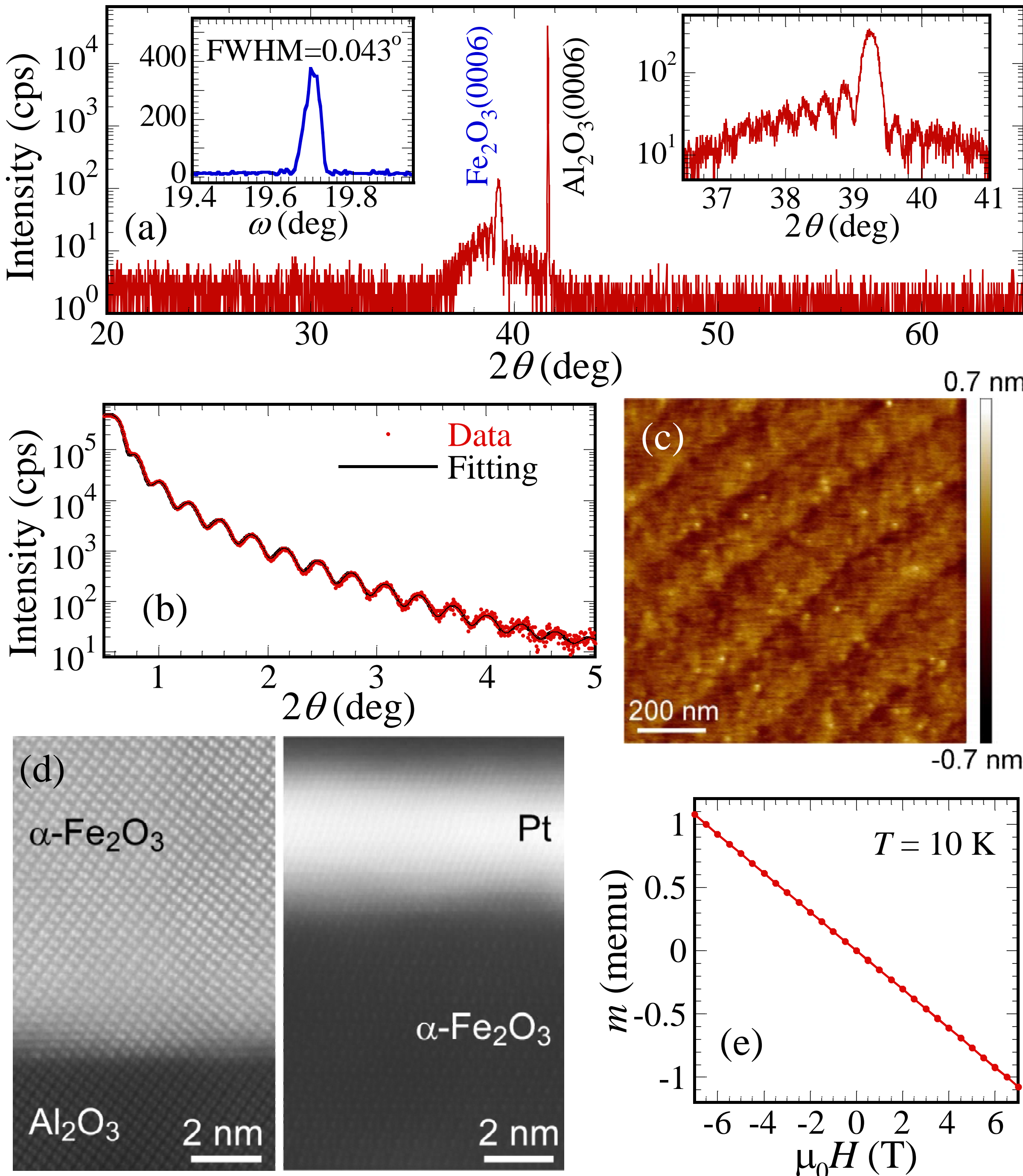

**Figure 1.**

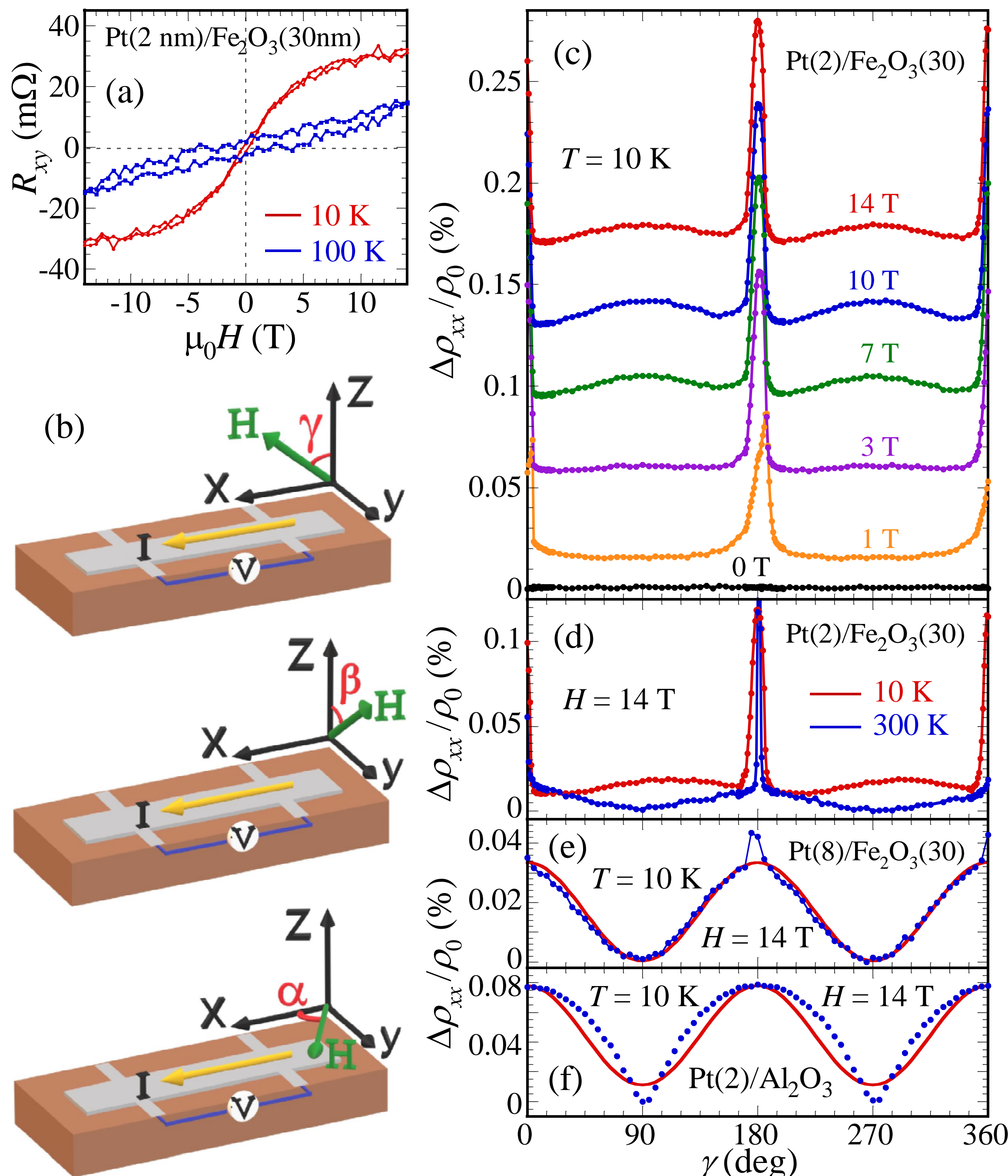

**Figure 2.**

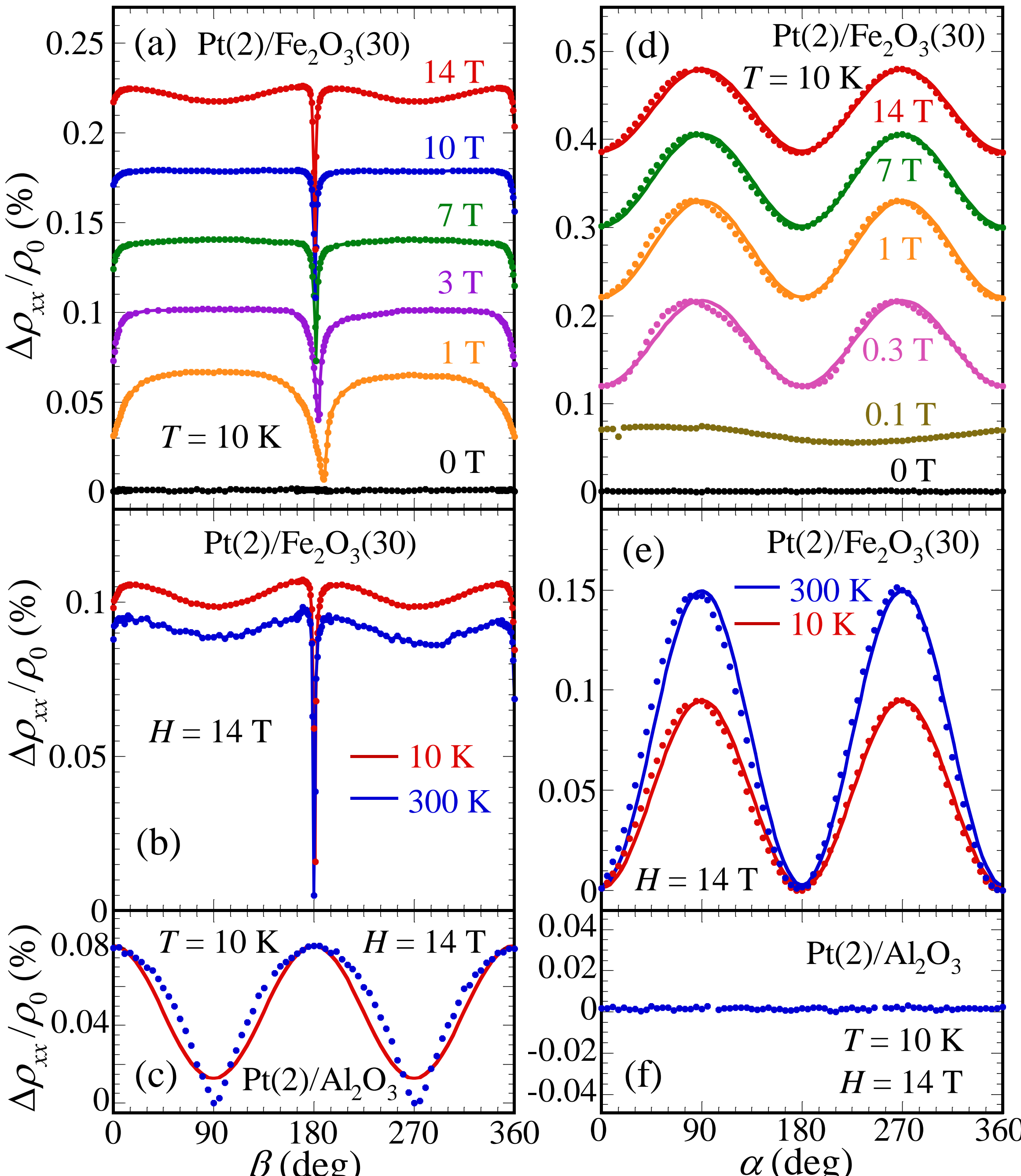

**Figure 3.**

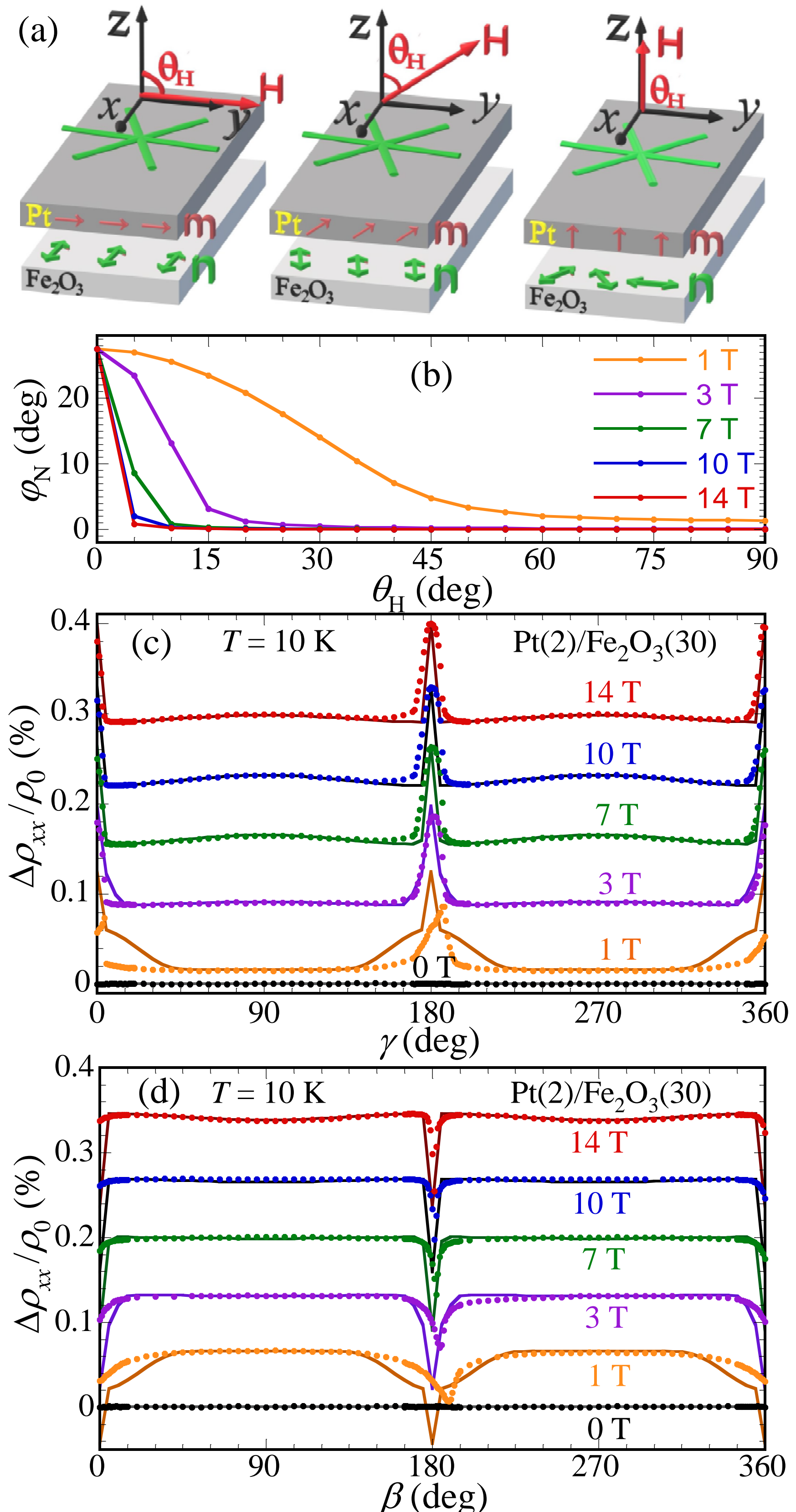

**Figure 4.**